\documentclass[12pt]{article}
\usepackage{epsfig}
\title{On the definition of velocity in doubly special relativity theories }
\author{Piotr Kosi\'nski\thanks{supported by KBN grant 5 P03B05620} , 
 Pawe{\l} Ma\'slanka$^*$ \\
Department of Theoretical Physics II \\
University of {\L}\'od\'z \\
Pomorska 149/153, 90 - 236 {\L}\'od\'z/Poland.}
\date{}

\begin{document}
\maketitle
\begin{abstract}
We discuss the definition of particle velocity in doubly relativity theories. The general formula relating
velocity and fourmomentum of particle is given.
\end{abstract}

\newpage

Recently there appeared a number of papers \cite{b1} $\div $\ \cite{b20} devoted to the so called doubly special relativity
theories (DSR). These are modifications of special relativity in which some particular value of energy/momentum
joins the speed of light as an invariant. It has been argued that there is a sound physical basis for such a modification.
However, up to now, DSR exists as a project rather than a complete theory. In particular, there is a controversy concerning 
the many-particle sector. In spite of that the following assumptions seem to be common to all versions of DSR theories:\\
(i) the momentum space is commutative; \\
(ii) the symmetry group \underline{is} the Lorentz group; 
 however, it can act in momentum space nonlinearly; \\
(iii)  rotation subgroup acts in the standard way, i.e. the symmetry linearizes on rotation subgroup and the fourmomentum 
transforms as a triplet plus singlet.

In the present note we intend to touch one point concerning the definition of velocity in DSR. There has been some
 discussion in the literature concerning the proper definition of velocity \cite{b19} or even there were claims
\cite{b21} that this problem points to a serious obstacle of the theory. This is in spite of the fact that the
satisfying solution has been already given more or less explicitly \cite{b6} \cite{b15}. Below we discuss this problem
in more general terms showing that there exists a consistent definition which can be applied in all versions
of DSR theories respecting the assumptions (i) $\div$\ (iii).

The main obvious demand is that the velocity is a property of reference frame rather than of a particular object.
In order to satisfy this demand we can make use of the fact that the group structure of space-time transformations
remains the same as in Einstein's theory; then, velocity can be viewed as a parameter of Lorentz transformations.

We shall consider here the massive case, i.e. we assume that the set of allowed values of fourmomentum contains
the nontrivial element invariant under rotations. In order to characterize the action of Lorentz group in momentum
space let us remind that the classification of nonlinear realizations of groups is well known \cite{b22}.
Strictly speaking the theory presented in \cite{b22} applies to the compact group case. However, it can be adopted in 
the massive case considered here. This is because in the massive case there exists a point in momentum space which 
is invariant under the action of compact subgroup (rotations) and we are looking for realizations linearizing on this 
subgroup. The main conclusion coming from Ref.\cite{b22} is that there is basically  one realization parametrized by the 
elements $\Lambda^i_0$\ of the Lorentz group; it reduces to the triplet on rotation subgroup. By adding a singlet one
can linearize the action of Lorentz group, the additional constraint, necessary in order to keep the same number of 
independent coordinates, is simply the mass-shell condition. All other realizations are obtained by a, in 
general nonlinear, change of variables. Therefore, the final conclusion is that a nonlinear action of Lorentz group
on momentum space, 
\begin{eqnarray}
&&p'^{\mu}=f^{\mu}(p;\Lambda ) \label{w1} \\
&&f{^\mu}(f(p;\Lambda ),\Lambda ')=f^{\mu}(p;\Lambda '\Lambda )\nonumber 
\end{eqnarray}
has a form
\begin{eqnarray}
f^{\mu}(p;\Lambda )=F^{\mu}(\Lambda F^{-1}(p)) \label{w2}
\end{eqnarray}
where
\begin{eqnarray}
q^{\mu}\rightarrow p^{\mu}=F^{\mu}(q) \label{w3}
\end{eqnarray}
is some mapping from the standard fourmomentum space to the actual one. In general, this mapping transforms
 invertibly part of the standard energy-momentum space (for example, the interior of light cone) onto some sector of
modified space. This is because we want to keep some finite energy / momentum Lorentz invariant while for the
standard action of Lorentz group only infinity of energy-momentum space is such an invariant.

In the standard case the mass shell $\Sigma_m $\ is isomorphic to the coset space $SO(3,1)/SO(3)$\ and $SO(3)$\ is
a stability subgroup of $r=(m,\vec{0})$. Due to the fact that the map $F$\ preserves the linear action of $SO(3)$, 
$k^{\mu}=F^{\mu}(r)$\ takes the same form, $k=(M,\vec{0})$. Let us call $\tilde{\Sigma}_M$\ 
the image of $\Sigma_m$\ under $F$. Any
fourmomentum $p$\ belonging to $\tilde{\Sigma}_M$\ can be obtained from $k$\ by the action of some element
of Lorentz group
\begin{eqnarray}
p^{\mu}=f^{\mu}(k;\Lambda )\label{w4}
\end{eqnarray}

In order to define $\Lambda $\ in unique way we demand it to be a boost,
\begin{eqnarray}
\Lambda =B(\vec{v}) \label{w5}
\end{eqnarray}
corresponding to the velocity $\vec{v}$; the uniqueness follows from the fact that two Lorentz transformations giving rise
to the same $p$\ must differ by an element of stability subgroup of $k$, i.e. the rotation. Let us note
that the decomposition of an orbitrary element $\Lambda $\ of Lorentz group into the boots and rotation reads
\begin{eqnarray}
\Lambda =B\cdot R \label{w6}
\end{eqnarray}
where 
\begin{eqnarray}
B^{\mu}_{\;\nu}=\left\{
\begin{array}{ccl}
&&\Lambda^{\mu}_{\;0}\; ,\;\;\;\;\;\;\;\;\;\;\;\;  \nu =0 \\
&&-\Lambda_i^{\;0} \;,\;\;\;\;\;\;\;\;\;\;\;\mu =0,\;\nu=i \\
&&\delta^i_{\;j} \;-\frac{\Lambda^i_{\;0}\Lambda_j^{\;0}}{1+\Lambda^0_{\;0}},\;\;\;\;\;\;\;\mu =i,\;\nu =j
\end{array}
\right.  \label{w7}
\end{eqnarray}
and 
\begin{eqnarray}
R^{\mu}_{\;\nu}=\left\{
\begin{array}{ccl}
&&\delta^{\mu}_{\;\nu}\; ,\;\;\;\;\;\;\;  \mu =0\;or \;\nu =0 \\
&&\Lambda_{\;j}^{i}\;-\frac{\Lambda^i_{\;0}\Lambda^0_{\;j}}{1+\Lambda^0_{\;0}},\;\mu =i,\;\nu =j
\end{array}
\right.  \label{w8}
\end{eqnarray}
We arrived at the following relation 
\begin{eqnarray}
p^{\mu}=f^{\mu}(k;B(\vec{v}))\label{w9}
\end{eqnarray}
We define $\vec{v}$\ to be the velocity of the particle characterized by the fourmomentum $p$. It is easily
seen that this definition gives the proper addition law; indeed
\begin{eqnarray}
&&p'^{\mu}=f^{\mu}(p;\;B(\vec{v}\;'))=f^{\mu}(f(k;\;B(\vec{v}));\;B(\vec{v}\;'))= \nonumber \\
&&\;\;\;\;=f^{\mu}(k;\;B(\vec{v}\;')B(\vec{v}))=f^{\mu}(k;\;B(\vec{v}\oplus \vec{v})R)=\nonumber \\
&&\;\;\;\;=f^{\mu}(k;\;B(\vec{v}\;' \oplus \vec{v}))\label{w10}
\end{eqnarray}

where $\vec{v}\;' \oplus \vec{v}$\ denotes Einstein addition law for velocities.

By specifying the form of $F$\ one can give an explicite formula for the relation between fourmomentum $p$\ and
velocity $\vec{v}$. The general form of $F$\ reads
\begin{eqnarray}
&&p^0=F^0(q)=\tilde{H}(q^0,\;\vec{q}^{\;\;2})\equiv H(q^0,\;m^2)\nonumber \\
&&p^i=F^i(q)=\tilde{G}(q^0,\;\vec{q}^{\;\;2})q^i\equiv G(q^0,\;m^2)q^i \label{w11}
\end{eqnarray}
Using eqs. (\ref{w2}) and (\ref{w9}) one can easily show that
\begin{eqnarray}
\vec{v}=\frac{\partial H^{-1}(p^0)}{\partial p^0}(\frac{\partial }{\partial q^0}(q^0G(q^0))-\frac{m^2}{q^0}
\frac{\partial G}{\partial q^0})\mid_{_{q^0=H^{-1}(p^0)}}\cdot\frac{\partial p^0}{\partial \vec{p}}\label{w12}
\end{eqnarray}
where $\frac{\partial p^0}{\partial \vec{p}}$\ is calculated from the deformed mass-shell condition
\begin{eqnarray}
G^2(H^{-1}(p^0))(H^{-1}(p^0))^2-\vec{p}^{\;2}=m^2G^2(H^{-1}(p^0))\label{w13}
\end{eqnarray}

Let us use our formula in some particular cases. First, let us consider the $\kappa $-Poincare case.
The general form of the mapping (\ref{w11}) has been given in Ref.\cite{b23} and reads
\begin{eqnarray}
&&H(q^0,\;m^2)=\kappa \ln (\frac{q^0+C(m^2)}{C(m^2)-A(m^2)}) \nonumber \\
&&G(q^0,\;m^2)=\frac{\kappa}{q^0+C(m^2)}\label{w14}
\end{eqnarray}
where $A(m^2)$\ and $C(m^2)$\ obey
\begin{eqnarray}
A^2(m^2)-2A(m^2)C(m^2)+m^2=0 \label{w15}
\end{eqnarray}

The partial derivative $\frac{\partial p^0}{\partial \vec{p}}$\ is calculated from the deformed mass-shell relation
\begin{eqnarray}
\frac{2A}{C-A}=\frac{1}{\kappa^2}(4\kappa^2sh^2(\frac{p^0}{2\kappa})-e^{p^0/_{\kappa}}\vec{p}^{\;2})\label{w16}
\end{eqnarray}
Eq. (\ref{w12}) implies in this case 
\begin{eqnarray}
\vec{v}=\frac{2\vec{p}}{\kappa (1-e^{-2p^0/_{\kappa}}+\frac{\vec{p}^2}{\kappa^2})} \label{w17}
\end{eqnarray}

which coincides with the right group velocity defined in Ref \cite{b15}. 

As a second example consider the Magueijo-Smolin version of DSR \cite{b6}. 
The general form of the mapping (\ref{w11}) is now
\begin{eqnarray}
&&H(q^0,\;m^2)=\frac{\kappa q^0}{q^0+C(m^2)} \nonumber \\
&&G(q^0,\;m^2)=\frac{\kappa}{q^0+C(m^2)}\label{w18}
\end{eqnarray}
while the deformed mass-shell condition reads
\begin{eqnarray}
\frac{(p^0)^2-\vec{p}^{\;2}}{(1-\frac{p^0}{\kappa})^2}=\frac{\kappa^2m^2}{C^2(m^2)}\label{w19}
\end{eqnarray}

A simple computation gives  
\begin{eqnarray}
\vec{v}=\frac{\vec{p}}{{p^0}}\label{w20}
\end{eqnarray}
again in accordance with the original result \cite{b6}.

Let us note that in both cases $\vec{v} \ne \;\frac{\partial p^0}{\partial \vec{p}}$. This is related to the fact that
one should use a deformed hamiltionan formalism \cite{b24}.

\end{document}